# OPERATING EXPERIENCE WITH ELECTRON CLOUD CLEARING ELECTRODES AT DAΦNE

M. Zobov, D. Alesini, A. Drago, A. Gallo, S. Guiducci, C. Milardi, A. Stella, INFN LNF, Frascati;
S. De Santis, LBNL, Berkeley; T. Demma, LAL, Orsay; P. Raimondi, ESRF, Grenoble

*Abstract*
During the current run of an electron-positron collider DAΦNE special electrodes for electron cloud suppression have been inserted in all dipole and wiggler magnets of the positron ring. In this paper we discuss the impact of these electrodes on beam dynamics and overall collider performance. In particular we report results of measurements such as e-cloud instabilities growth rate, transverse beam size variation, tune shifts along the bunch train etc. with the electrodes switched on and off that clearly indicate the effectiveness of the electrodes for e-cloud suppression.

## INTRODUCTION

DAΦNE is the electron-positron collider working at the energy of 1 GeV in the centre-of-mass, the energy of the Φ-resonance [1]. The accelerator complex consists of a full energy Linac, an Accumulator/Damping ring and two main rings intersecting in two interaction regions. Typically we use a single interaction point (IP) and the beams are separated at the second IP in one way or another [2]. Table 1 shows main collider parameters.

Table 1. DAΦNE main parameters

| Energy per beam, MeV | 510 |
|---|---|
| Machine length, m | 97 |
| Max. beam current, A (KLOE run) | 2.5 (e-), 1.4 (e+) |
| N of colliding bunches | 100-110 |
| RF frequency, MHz | 368.67 |
| RF voltage, kV | 120-220 |
| Harmonic number | 120 |
| Bunch spacing, ns | 2.7 |
| Peak luminosity, $cm^{-2}s^{-1}$ (SIDDHARTA run) | $4.5 \cdot 10^{32}$ |

The maximum luminosity was achieved in collisions with the novel crab waist scheme [3]. Its best value of $4.5 \times 10^{32}$ $cm^{-2}s^{-1}$ is by two orders of magnitude higher than the luminosity of Novosibirsk VEPP-2M collider [4] that was working at the same energy. This result was achieved by storing beams with high currents distributed over many colliding bunches. The maximum electron beam current was 2.5 A, i.e. the record value for the electron beam current ever stored in the modern colliders and synchrotron light sources. However, due to the electron cloud effects we have not been able to exceed 1.4 A in the positron ring. In addition to this current limitation we have been suffering from other harmful e-cloud effects affecting the collider luminosity performance such as anomalous pressure rise, vertical beam size increase, tune spread along the bunch train etc. [5].

In order to cope with the strong e-cloud instabilities powerful bunch-by-bunch feedback systems have been used [6]. In addition, solenoids were installed in order to prevent the e-cloud build-up in the straight sections. However, these measures have not solved completely the problems created by the e-cloud. In particular, the strong horizontal instability still remains the worst trouble for us. Numerical simulations and experimental observations have shown that this kind of instability is triggered by the e-cloud pattern created in the wiggler and dipole magnets [7]. So it has been decided to insert special metallic electrodes to suppress the e-cloud in the dipoles and the wigglers.

According to the dedicated numerical simulations application of the DC voltage of 200-500 V at the electrodes should decrease the electron cloud density by about 2 orders of magnitude. This is in accordance with experimental measurements of the electron cloud density suppression in CERN PS [8], CESR [9] and KEKB [10] where single test clearing electrodes were installed in wiggler or dipole sections.

Differently from [8-10] in DAΦNE the e-cloud clearing electrodes have been inserted in all dipole and wigglers magnets so that their effectiveness can be verified in collider operational conditions looking at the tune shift and tune spread in the bunch trains, instabilities growth rate, transverse beam sizes, luminosity and overall collider performance. The electrode design is also different from that used in [8-10].

In this paper we describe the design of the clearing electrodes and discuss results of different experimental measurements and observations proving the effectiveness of the electrodes for e-cloud suppression and improving the collider performance.

## CLEARING ELECTRODES

### Design and Installation

It was decided to install the electron cloud clearing electrodes in place without opening the DAΦNE vacuum chamber by inserting the electrodes through lateral vacuum pump ports. This was one of the reasons why we use the rigid metallic electrodes. Such electrodes are also technologically simpler and cheaper than those used in [8-

10] made of a very thin highly resistive layer deposited on a thin ceramic substrate.

The pictures of the electrodes inserted in the dipole and wiggler chambers are shown in Fig. 1. The dipole electrodes have a length of 1.4 or 1.6 m depending on the considered collider arc, while the wiggler ones are 1.4 m long. They have a width of 50 mm, thickness of 1.5 mm and their distance from the chamber is about 0.5 mm. This distance is guaranteed by special ceramic supports made in SHAPAL (Fig. 2) and distributed along the electrodes. This ceramic material is also thermo-conducting in order to partially dissipate the power released from the beam to the electrode through the vacuum chamber coupling impedance. Moreover, the supports have been designed to minimize their contribution to the coupling impedance and to simultaneously sustain the strip. The mechanical drawing of a dipole-wiggler arc with the electrodes is shown in Fig. 3. The distance of the electrode from the beam axis is 8 mm in the wigglers and 25 mm in the dipoles. Unfortunately, due to mechanical installation constraints the wiggler electrodes do not cover the total length of the wigglers that are 2.1 m long (to be compared with 1.4 long electrodes).

The electrodes have been connected to external dc voltage generators modifying the existing BPM flanges as shown in Fig. 4.

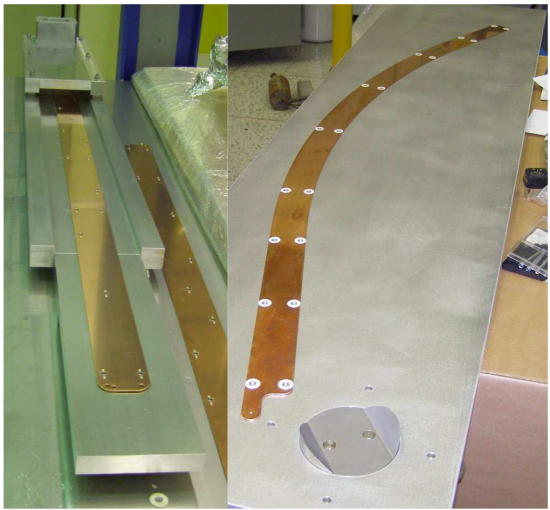

Fig. 1: Pictures of the electrodes inserted in the dipole (right) and wiggler (left) chambers.

The electrodes have been inserted in the vacuum chamber during January-May 2010 shutdown. The picture of an installed electrode is given in Fig. 5. The electrodes have been inserted in the machine using special plastic supports that allowed inserting the electrodes in the chamber without damaging the chamber and the electrodes themselves. Before and after their installation the electrodes have been tested applying a dc voltage of about 400 V (in air) to check the correct installation and reliability of the connections. Measurements with a Network Analyzer have also been done and they will be illustrated in the next section.

A low pass-band RC filter has been inserted between the feed-through and the dc generator in order to decouple the dc generator from the beam induced signal at high frequency.

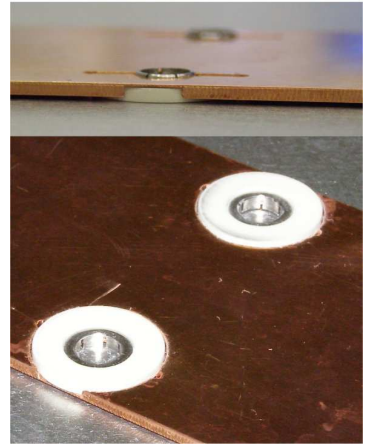

Fig. 2: SHAPAL supports for the electrodes.

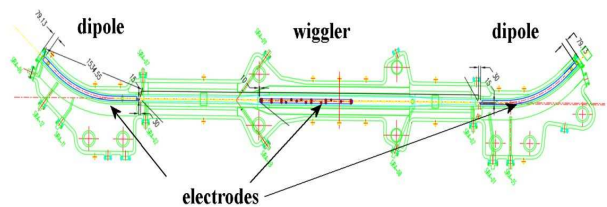

Fig. 3: Mechanical drawing of a complete arc with the electrodes.

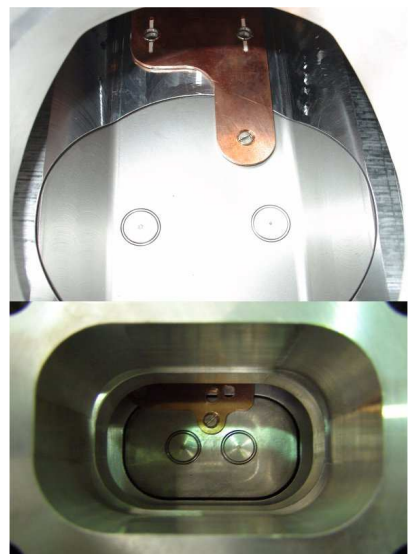

Fig. 4: Detail of the electrodes output connection.

*Electrode Coupling Impedance*

The electrode installation in the ring was a risky operation from the beam impedance point of view. In the past we had a negative experience with the so-called "invisible" ion clearing electrodes that contributed almost a half of the electron ring impedance and at some point we had to remove them [11]. So much attention has been dedicated to the impedance studies. Here we consider the most critical case of the wiggler electrodes.

The total electrode coupling impedance consists of the resistive wall impedance due the finite conductivity of the electrode and the strip-line impedance since a real strip-line is created between the electrode and the vacuum chamber walls. It has been estimated that for the wiggler electrode the resistive wall contribution alone would result in the temperature rise up to 50-55 degrees under vacuum [12]. In turn, the strip-line impedance depends much on the external matching conditions. Even for a hypothetical case of a perfectly matched electrode the loss factor would be by a factor 3 higher than that of the resistive wall. The perfect matching is almost impossible and one could expect even higher beam losses. So, in order to keep the situation under control it has been decided to: a) mismatch the electrode intentionally to have the resonances very narrow and b) to choose the electrode length in such a way to have the powerful RF harmonic frequencies just between the dangerous resonance lines. The overall broad-band impedance was reduced by decreasing the "electrode-wall" characteristic impedance by making the gap smaller and the electrode wider.

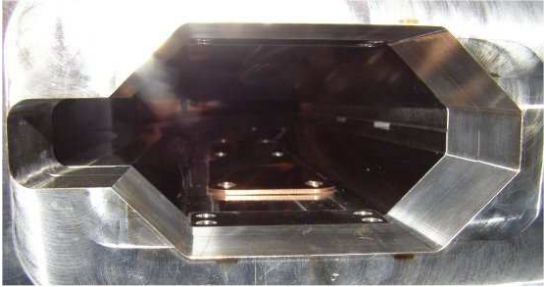

Fig. 5: Installed electrode in the dipole vacuum chamber.

In order to obtain wake fields and impedance of the electrode we have used the code GdfidL [13]. The simulations have been performed for real electrode geometry with one feed-through placed at one of the electrode ends and including the dielectric supports. We have traced the wake behind the exciting bunch over 50 m and the resulting impedance was obtained by applying the Fourier transform. The real part of the electrode impedance is shown in Fig. 6.

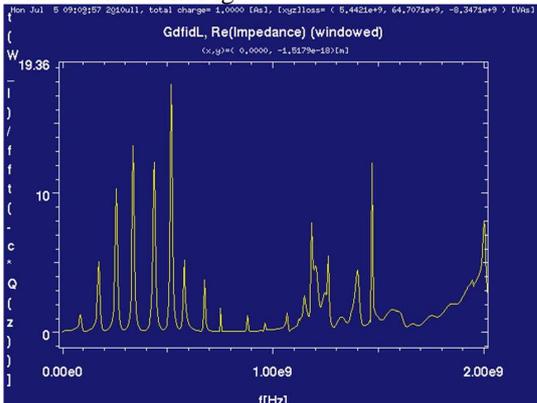

Fig. 6: Real part of the electrode coupling impedance

RF measurements with a network analyzer have been performed before and after the electrode installation. We have done two types of measurements: reflection coefficient at the feed-through port and transmission coefficient between one BPM near to the strip and the feed-through. In both cases it was possible to measure the resonant frequencies of the strip modes and, especially in the second case, it was possible to measure the resonant frequencies also with the presence of the RC low pass-band filter. As an example the first type of measurements is reported in Fig.7. As we can see comparing Fig.6 and Fig. 7 the resonance line pattern is similar in the simulations and the measurements. Moreover, as shown in Fig.7 the powerful RF harmonics do not couple to the resonance lines thus helping to minimize power losses in multibunch operations.

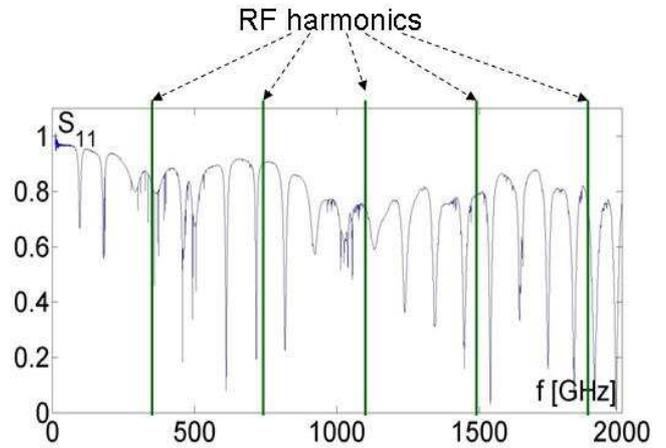

Fig. 7: Reflection coefficient at the feed-through port.

Nevertheless the lost power is not negligible and can result in excessive heating of the electrode. In order to prevent this possible damage, the electrode supports are made of thermo-conducting dielectric material (SHAPAL) thus providing heat transfer from the electrode to the vacuum chamber.

The estimated low frequency broad-band impedance of the electrode $Z/n$ is about 0.005 Ω, which is substantially smaller with respect to the impedance of the ion clearing electrodes removed from the wiggler sections of the electron ring [11], and should be a small contribution to the total ring impedance. Indeed, this has been confirmed by measuring bunch lengthening in both the electron and the positron rings and comparing the results. We have not found any relevant differences.

*Effectiveness Evaluation*

In order to evaluate the effectiveness of e-cloud suppression by the electrodes we have used the code ECLOUD [14] modified to include the effect of a vertical electric field on the electron cloud dynamics. The simulations have been performed with the following simplifying assumptions:

- The vacuum chamber is considered elliptic.

- The magnetic field is assumed to be uniform and equal to the maximum value in the wiggler.
- The electric field is vertical and uniform in the region of the chamber occupied by the electrodes and zero outside this region. The value of the field is obtained from: E=V/L (L is the vertical size of the chamber and V the potential of the electrode with respect to the chamber).
- For these simulations the SEY of aluminum has been used both for the chamber and electrode surface.

Figure 8 shows the electron cloud density evolution for different values of the electrode voltage for parameters reported in Table 2. The listed beam parameters correspond to the collider operation with the positron current of 800 mA.

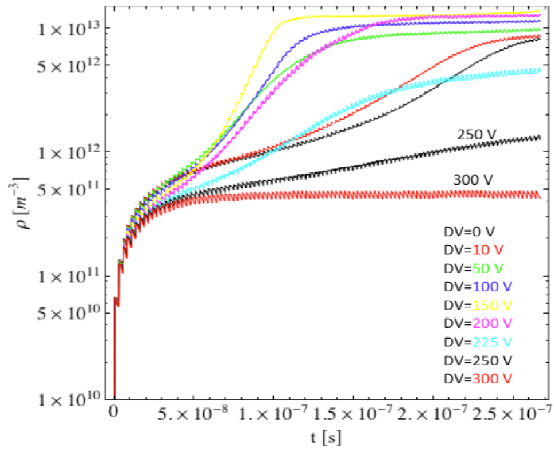

Fig 8: Evolution of the beam chamber averaged cloud density as computed by ECLOUD. Different colours correspond to different values of the electrode voltage as indicated in the figure.

Table 2: Input parameters for ECLOUD simulations.

| Parameters | Value |
| --- | --- |
| Bunch population | $1.68 \times 10^{10}$ |
| Number of bunches | 100 |
| Bunch spacing [m] | 0.8 |
| Bunch length [mm] | 12.3 |
| Bunch horizontal size [mm] | 1.08 |
| Bunch vertical size [mm] | 0.05 |
| Hor./vert. chamber sizes [cm] | 12/2 |
| Primary electron rate | 0.0088 |
| Photon Reflectivity | 100% (uniform) |
| Maximum SEY | 1.9 |
| Energy at max. SEY [eV] | 250 |
| Vertical magnetic field [T] | 1.64 |

It is clearly observed a non-monotonic dependence of the saturation density on the electrode voltage. In particular the maximum value is reached around V=200 V. For higher voltages the density sharply decreases. We see that already at 300 V it is reduced by about two orders of magnitude. The same behavior is observed for different values of the beam current, as shown in Figure 9, and is in qualitative agreement with measured currents absorbed by the clearing electrodes (see the last paragraph of the paper).

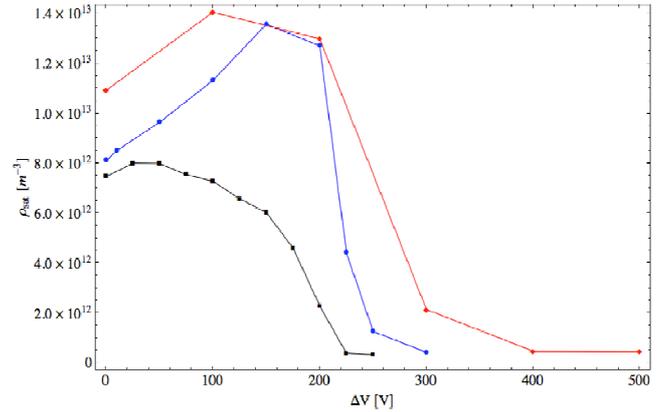

Fig 9: Electron cloud density at saturation (i.e. at the end of bunch train) as a function of the electrode voltage as computed by ECLOUD. Different colours correspond to different beam currents: 500 mA-Black; 800 mA-Blue; 1000 mA-Red.

Such a complicated e-cloud density behavior as a function of the electrode voltage is now under study.

## EXPERIMENTAL MEASUREMENTS

Several experimental measurements have been performed to check the effectiveness of the electrodes to suppress the electron cloud. Their effect on the positron beam has been observed using a synchrotron light monitor, an FFT spectrum analyzer, and the bunch-by-bunch horizontal and vertical feedback systems. Besides, we can extract some useful information on the e-cloud suppression analysing frequency shifts of HOMs trapped in the vacuum chamber and measuring the current absorbed by the electrodes.

### Tune Shift and Tune Spread

The first and most obvious measurement is the measurement of the "average" betatron tune shift by switching on and off the electrodes. The horizontal tune shift measurements with electrodes on and off are given in Fig. 10 for a 550 mA positron beam (not in collision). The image is taken from a spectrum analyser (Tektronix RSA3303A) connected to a button pickup. The frequency shift of the horizontal tune line switching off all electrodes is ≈20 kHz which correspond to a difference in the horizontal tune of ≈0.0065. The betatron tune is shifted in the positive direction while switching off the electrodes. This is a clear indication that the electron cloud density is reduced.

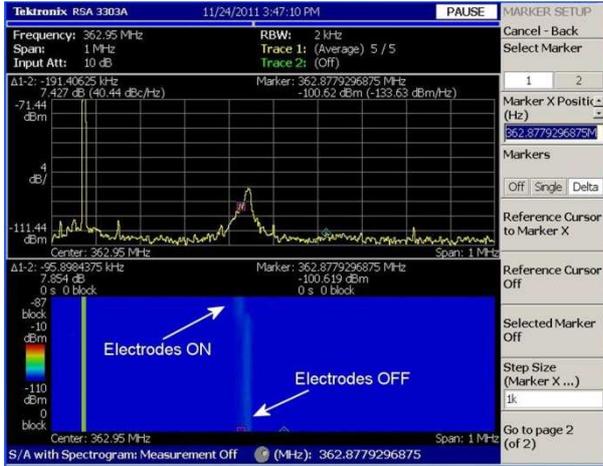

Fig. 10: Horizontal tune shift measured with 550 mA of positron beam. The image is taken from the spectrum analyser connected to one button pickup.

More sophisticated tune measurement has been performed using capabilities of the DAΦNE bunch-by-bunch feedbacks [15]. Off-line analysis of the signals acquired by the bunch-by-bunch transverse feedbacks allows measuring the fractional tunes of each bunch along the bunch batch. This provides a very powerful tool to observe the tune shift modulation due to the electron cloud along the bunch train.

Fig. 11 and Fig. 12 show the measured tune shifts as a function of the bunch number in the bunch train for the horizontal and vertical planes, respectively. These measurements were performed by turning on and off all four wigglers electrodes and two (out eight) dipole electrodes. The bunch train was composed of 100 consecutive bunches separated by 20 bunch gaps.

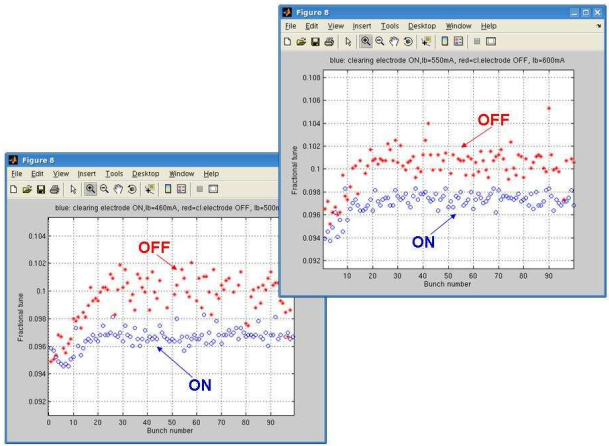

Fig. 11: Two examples of measurements of horizontal fractional tune as a function of bunch number (blue points: electrodes ON, red curve: electrodes OFF).

In these figures we can observe the typical tune modulation along the train induced by the electron cloud. The fractional tunes increase in the first part of the train where the e-cloud density grows up and reaches a steady state regime after ≈20 bunches. In the horizontal plane the tune spread between the head and the tail of the train is about 0.006-0.008 with electrodes off (red points). It reduces by a factor 2-3 when the electrodes are switched on (blue points). We see that the electrodes are very effective but they do not cancel completely the tune spread. We attribute this to the fact that the electrodes in the wigglers cover only 67% of their total length. Besides, not all the dipole electrodes were switched on during the measurements. In turn, as it is seen in Fig. 12, the vertical tune spread is notably smaller than the horizontal one and the electrodes almost completely cancel it.

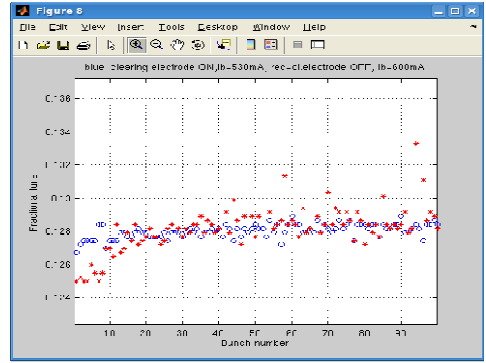

Fig. 12: Vertical fractional tune as a function of bunch number (blue points: electrodes ON, red curve: electrodes OFF).

*Instability growth rate*

Yet another useful measurement with the feedback system is the measurement of instability growth rates [16]. Typically the feedback is switched off for a short period of time and the instability amplitude is acquired turn-by-turn bunch-by-bunch. Then the modal analysis is performed and the instability growth rate is calculated, see Fig. 13.

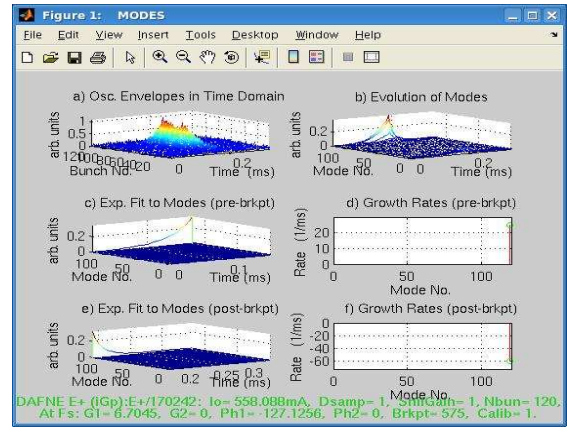

Fig. 13. Example of growth rate measurements.

Here you can see that the unstable mode is the mode -1, similar to the resistive wall instability. In the past it has been predicted by simulation that exactly this mode becomes unstable due to the electron cloud created in the dipole and wiggler magnets and having the shape of two vertical parallel stripes [7].

Figure 14 summarizes the growth rate measurement results where the growth rates (in ms$^{-1}$) are plotted as a function of the positron beam current for different electrode voltages. With electrodes off the growth rate at 650 mA exceeds 50 ms$^{-1}$ and the measurements above this current become quite difficult since the beam is strongly unstable. With electrodes on these growth rates are strongly reduced and it is possible to store a higher stable current. In the future we plan to perform more measurements at higher electrode voltages and for different bunch patterns.

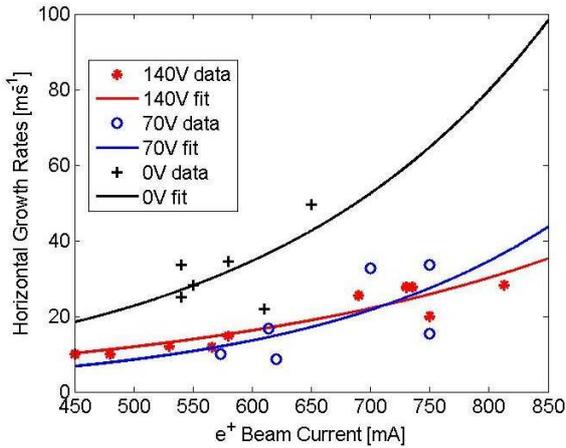

Fig.14: Growth rates of the horizontal instability.

*Transverse beam size*

Looking at the synchrotron light monitor we can observe the vertical beam size enlargement by gradually turning off the electrodes, one after another, as shown in Fig. 15. As it is seen, the vertical size increases from about 110 μm with electrodes on to more than 145 μm with the electrodes off. Also in this case the beams were not colliding.

There is some observable horizontal size reduction and one can think of coupling introduced by the electrodes. However, analysis has shown that the coupling alone cannot explain such a big vertical size increase. What is even more important, we see also the corresponding luminosity increase with the electrodes on when the beams are in collision.

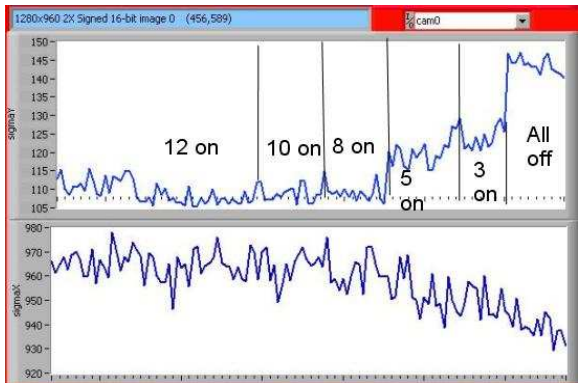

Fig. 15: Beam dimension (in μm) at the SLM turning off, progressively, all electrodes.

*Frequency Shift of Vacuum Chamber HOMs*

The e-cloud plasma can interact with RF waves transmitted in the vacuum chamber changing the phase velocity of the waves. Such measurements have been successfully done on other machines [17]. A similar approach can be used in case of resonant waves in the vacuum chamber. Even in this case the e-cloud changes the electromagnetic properties of vacuum and this can result in a shift of the resonant frequencies of vacuum chamber trapped modes. In principle, from these shifts it is possible to evaluate the e-cloud density [18].

Resonant TE-like modes are trapped in the DAΦNE arcs and can be excited through button pickups. The lower modes have frequencies between 250 and 350 MHz. A first measurement of these resonant modes has been done at DAΦNE for several beam currents with the electrodes on and off [19]. The preliminary analysis of this data has given the following results: (a) all modes have a positive frequency shift as a function of the positron beam current. For 800 mA it is between 100 and 400 kHz depending on the modes we are considering; (b) for almost all the modes we can partially cancel the frequency shift switching on the electrodes; (c) the quality factor of the modes decreases with positron current. The fact that for some modes the shift does not depend on the electrode voltage could depend on the fact that these modes are localized in different places of the arc and also in regions not covered by electrodes. For instance the transmission coefficient between two button pickups in the arc chamber is given in Figs. 16a and 16b for two different modes. The mode of Fig. 16a has a positive frequency shift that does not change with electrode voltage while the mode in Fig. 16b corresponds to the case of a mode with frequency dependent to the electrode voltage. An identification of the resonant mode locations and further analysis of their coupling to the e-cloud pattern in the wigglers and dipoles are still in progress. These studies can be useful to evaluate the e-cloud densities in different parts of the vacuum chamber.

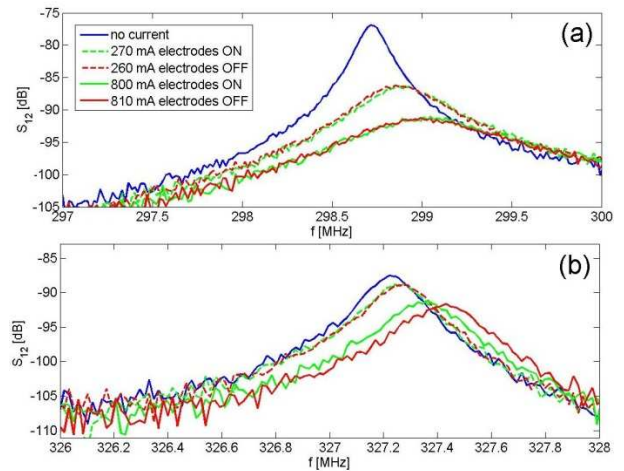

Fig. 16: Transmission coefficient between two button pickups in the arc chamber for two different resonant modes.

*Current Absorbed by Electrodes*

The voltage generators connected to the electrodes absorb the e-cloud electrons. In the present layout one voltage generator is connected to three electrodes of one arc (i.e. one wiggler and two dipoles). The current delivered by the generator has been measured as a function of the generator voltage for different beam currents. The result is given in Fig. 17.

Comparing the plots of Fig. 17 with those obtained by the simulations (see Fig. 9) we can see their qualitative agreement. In particular, we see that a) while increasing the generator voltage the e-cloud density initially increases, reaches its maximum and subsequently decreases; b) the density peak locations depend on the beam current and are shifted towards higher voltages for higher beam currents.

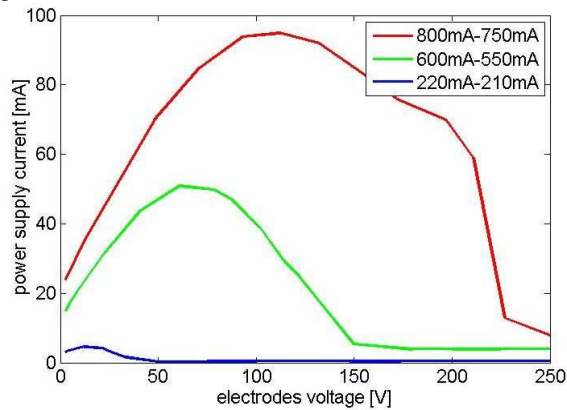

Fig.17: Current supplied by the dc voltage generator as a function of the applied voltage and beam current.

On the basis of the numerical predictions and the measurement results we can conclude that in order to store positron beam currents higher than 1A a voltage of the order of 250 V (presently available) is no longer adequate to completely absorb and suppress the e-cloud in DAΦNE. So the applied voltage has to be increased.

Another useful observation is that in order to prevent damaging the voltage generator due to the high absorbed e-cloud current and eventually the electrode from the bombardment by the e-cloud electrons it is worthwhile to apply negative voltages at the electrodes. The latter is particularly valid for the thin layers electrodes used in [8-10]. In DAΦNE operations we have found that the effectiveness of the electron cloud suppression does not depend on the voltage polarity while the absorbed current at -250 V reduces almost to zero. This is in agreement with similar observations made in Novosibirsk [20] and at CERN PS [8].

## CONCLUSIONS

Metallic e-cloud clearing electrodes have been inserted in all DAΦNE dipole and wiggler magnets and are now used in routine operations of the collider.

The electrodes are found to be very useful to reduce the strength of the positron beam horizontal instability; to decrease the betatron tune shift and tune spread inside bunch trains; to suppress the vertical beam size blow-up of the positron beam.

As a result, with the electrodes switched on it is possible to store higher positron beam current, to achieve higher luminosity and to have more stable overall collider performance.

## ACKNOWLEDGMENT

We would like to thank A. Battisti, V. Lollo and R. Sorchetti for the technical support in the electrode design and installation and O. Coiro for his help in voltage generator installation. The research leading to these results has received partial funding from the European Commission under the FP7 project HiLumi LHC, GA no. 284404, co-funded by the DoE, USA and KEK, Japan and by the European Commission FP7 Program EuCARD, WP11.2, Grant Agreement 227579.